\documentclass[twoside,12pt]{article}
\usepackage{epsfig}
\def\Journal#1#2#3#4{{#1} {#2} (#4) #3 }

\def\NPA{{\em Nucl. Phys.} A}

\def\PLB{{\em Phys. Lett.} B}

\def\PRC{{\em Phys. Rev.} C}

\newcommand{\be}{\begin{equation}}
\newcommand{\ee}{\end{equation}}
\newcommand{\bea}{\begin{eqnarray}}
\newcommand{\eea}{\end{eqnarray}}

\usepackage{graphicx}

\begin{document}
\title{\vspace{1cm} Statistical Model Predictions for p+p and Pb+Pb Collisions\\ at LHC}
\author{I.\ Kraus${}^{1,2}$, J.\ Cleymans${}^{1,3}$, H.\ Oeschler${}^{1}$, K.\ Redlich${}^{1,4}$, S Wheaton${}^{1,3}$\\
\\
$^1$ Institut f\"ur Kernphysik, Darmstadt University of Technology,\\ D-64289 Darmstadt, Germany \\
$^2$ Nikhef, Kruislaan 409, 1098 SJ Amsterdam, The Netherlands \\
$^3$ UCT-CERN Research Centre and Department  of  Physics,\\ University of Cape Town, Rondebosch 7701, South Africa \\
$^4$ Institute of Theoretical Physics, University of Wroc\l aw,\\ Pl-45204 Wroc\l aw, Poland}
\maketitle
\begin{abstract}
Particle production in p+p and central Pb+Pb collisions at LHC is discussed in the context of the statistical thermal model. For heavy-ion collisions, predictions of various particle ratios are presented. The sensitivity of several ratios on the temperature and the baryon chemical potential is studied in detail, and some of them, which are particularly appropriate to determine the chemical freeze-out point experimentally, are indicated.
Considering elementary interactions on the other hand, we focus on strangeness production and its possible suppression. Extrapolating the thermal parameters to LHC energy, we present predictions of the statistical model for particle yields in p+p collisions. We quantify the strangeness suppression by the correlation volume parameter and discuss its influence on particle production. We propose observables that can provide deeper insight into the mechanism of strangeness production and suppression at LHC.
\end{abstract}

\section{Introduction}
The large hadron collider (LHC) will deliver soon and for the first time p+p interactions at $\sqrt{s}$=~14~TeV and Pb+Pb collisions at $\sqrt{s_{nn}}$=~5.5~TeV. Hadron multiplicities are some of the very first measurements. Therefore it is useful to have theoretical predictions as benchmarks, still before first data become available. The statistical thermal model is one candidate to make such predictions since it successfully describes particle production at lower energies, e.g. Ref.~\cite{pbm}.

In Section~\ref{pb} predictions of the statistical model formulated in the grand-canonical ensemble are calculated and results for central Pb+Pb collisions are presented and discussed. We introduce the canonical formulation of strangeness conservation in Section~\ref{pp} together with a proposal to modify the treatment of the strange-particle phase-space. We investigate possible scenarios in p+p collisions at LHC energies and propose their possible experimental verification. We close with conclusions in Section~\ref{summary}.
%
\section{Particle production in Pb+Pb collisions} \label{pb}
Based on statistical model analyses of particle production at lower energies we first extrapolate the relevant model parameters to LHC energy and give some hadron ratios as examples. Then we quantify and discuss the sensitivity of different particle ratios to temperature and baryon chemical potential. Finally, we point out some which could be used to verify experimentally the freeze-out conditions.
\subsection{\it Extrapolation of thermal parameters and model predictions}
\begin{figure}[h]
\begin{center}
\begin{minipage}[b]{0.05\linewidth}
~
\end{minipage}\hfill
\begin{minipage}[b]{0.45\linewidth}
\centering
\includegraphics*[width=\linewidth]{figure_1a.eps}
~
\end{minipage}\hfill
\begin{minipage}[b]{0.05\linewidth}
~
\end{minipage}\hfill
\begin{minipage}[b]{0.4\linewidth}
\centering
\includegraphics*[width=\linewidth]{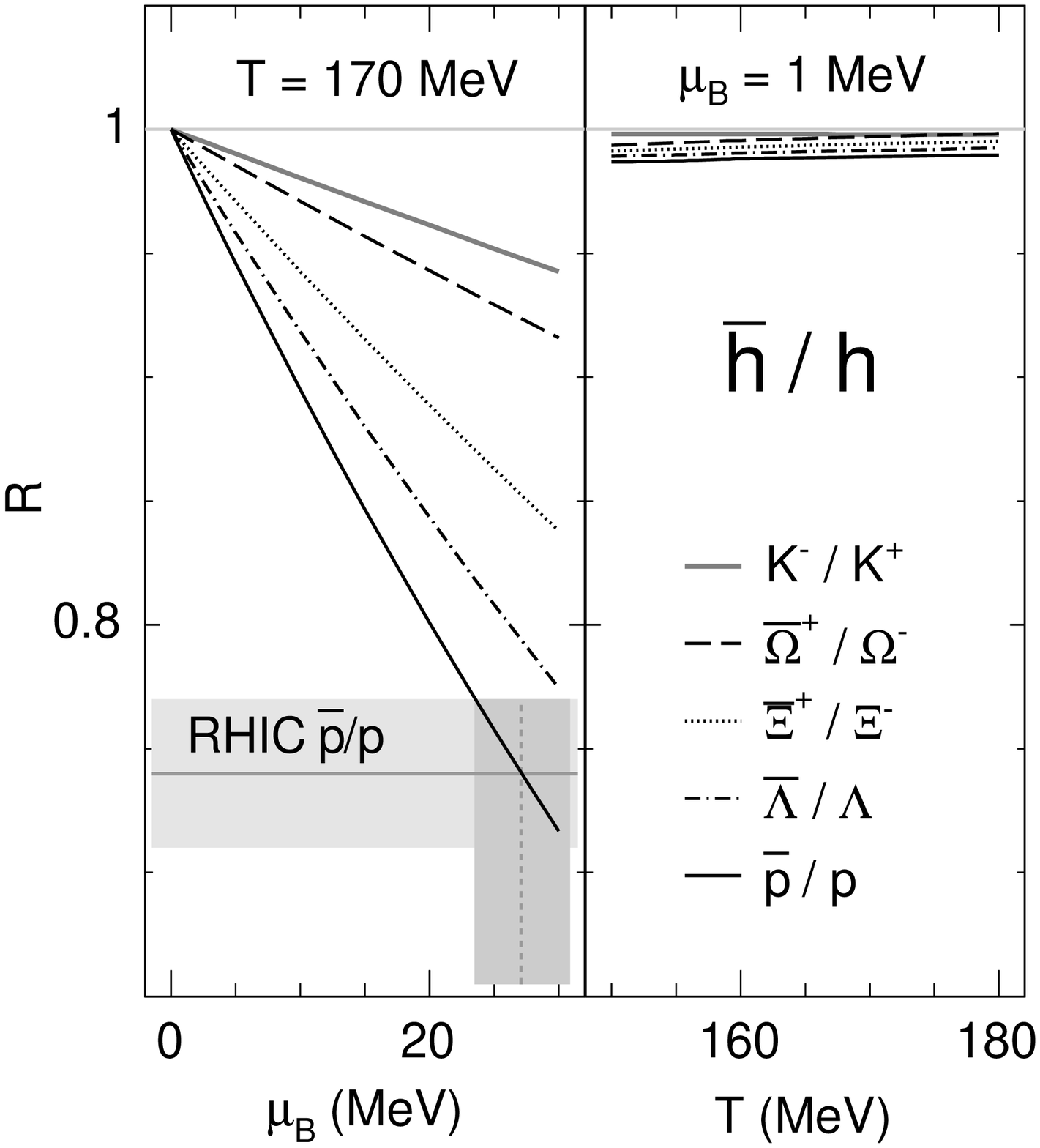}
\end{minipage}\hfill
\begin{minipage}[b]{0.05\linewidth}
~
\end{minipage}\hfill
\begin{minipage}[b]{16.5 cm}
\caption{\label{fig1ab}
Left: Energy dependence of the chemical freeze-out parameters $T$ and $\mu_B$. The points display results of statistical model analysis of data ranging from SIS to RHIC energies. The curve has been obtained using a parameterization discussed in Ref.~\cite{fo_curve}. The figure is taken from Ref.~\cite{fo_curve}.
\newline
Right: Antiparticle/particle ratios $R$ as a function of $\mu_B$ for $T$ = 170~MeV (left) and as a function of $T$ for $\mu_B$ = 1~MeV (right). The horizontal line at 1 is meant to guide the eye. The $\rm \bar{p}/p$ ratio (averaged over the data of the 4 RHIC experiments at $\sqrt{s_{NN}}$~=~200~GeV) is displayed (gray horizontal line) together with its statistical error (gray band). As illustrated, $\mu_B \approx \rm 27~MeV$ (dashed line) can be read off the Figure directly within the given accuracy (vertical gray band). See Ref.~\cite{lhc_pbpb} for details.}
\hfill
\end{minipage}
\end{center}
\end{figure}

As we have already seen in heavy-ion collisions at lower energies, the statistical model describes particle production successfully~\cite{kr,pbm}. Owing to the large system size and the large number of produced hadrons, the grand-canonical formulation is applicable. Consequently, there are only two parameters, the temperature $T$ and the baryon chemical potential $\mu_B$ required to quantify any particle ratio. The extracted pairs of thermal parameters, obtained from analyses of particle yields in Pb+Pb and Au+Au collisions at lower energies, are displayed in Fig.~\ref{fig1ab}. They feature a common behavior, which was vastly studied (see Ref.~\cite{fo_curve} and references therein); here we follow the parametrisation for the freeze-out curve given in Ref.~\cite{fo_curve} and extrapolate for predictions at $\sqrt{s_{nn}}$=~5.5~TeV the thermal parameters to $T$ = (170$\pm$5)~MeV and $\mu_B$~=~$\rm 1^{+4}_{-1}$ MeV.
%

Predictions of the statistical model for some particle ratios at LHC energy are listed in Table~\ref{Table1}.
The errors given in Table~\ref{Table1} reflect the systematic uncertainty in the extrapolation of the thermal parameters, see Ref.~\cite{lhc_pbpb} for a more detailed discussion.
\subsection{\it Experimental verification of model parameters}
\begin{figure}[h]
\begin{center}
\begin{minipage}[b]{0.05\linewidth}
~
\end{minipage}\hfill
\begin{minipage}[b]{0.45\linewidth}
\centering
\includegraphics*[width=\linewidth]{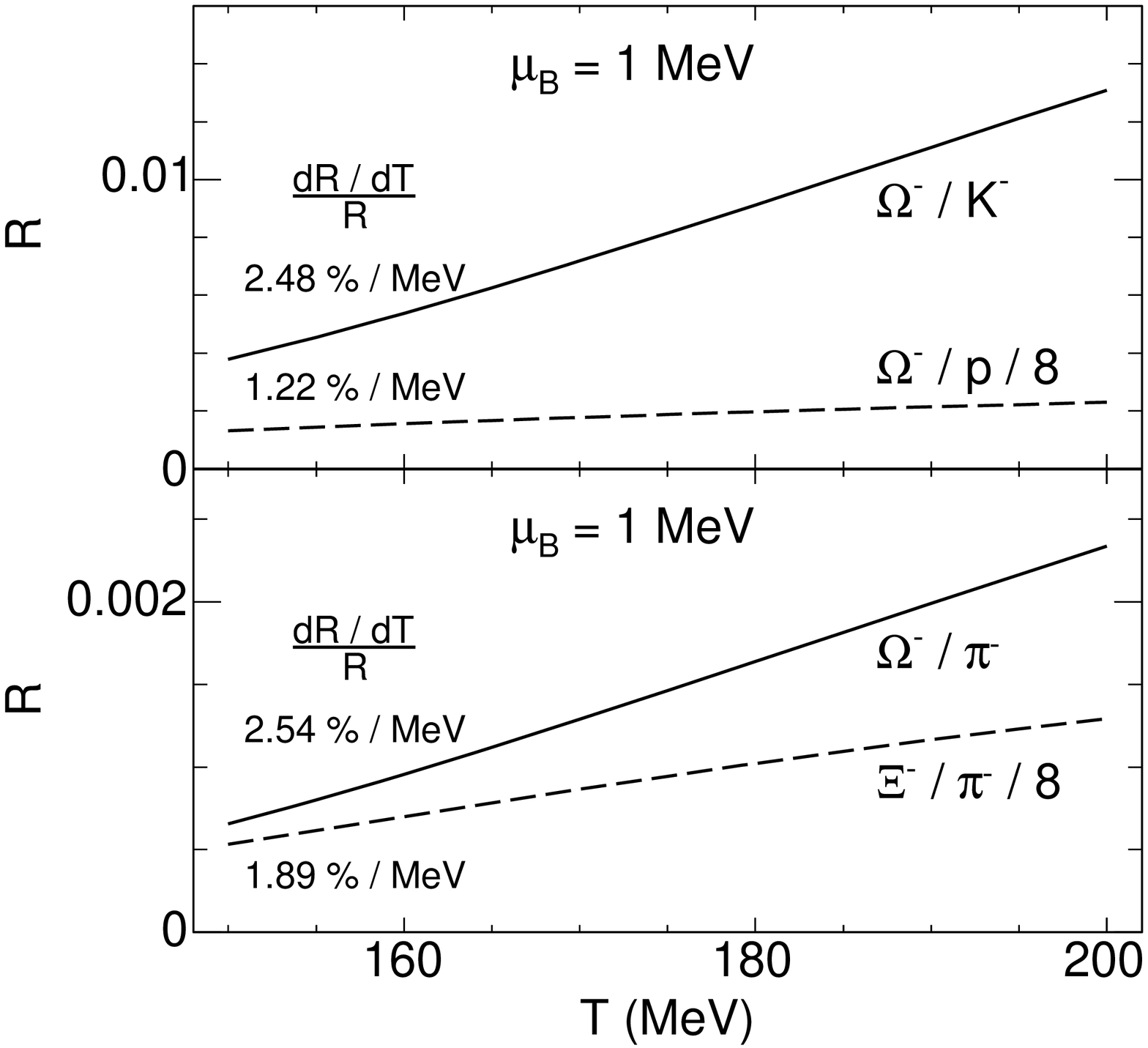}
~
\end{minipage}\hfill
\begin{minipage}[b]{0.05\linewidth}
~
\end{minipage}\hfill
\begin{minipage}[b]{0.39\linewidth}
\centering
\includegraphics*[width=\linewidth]{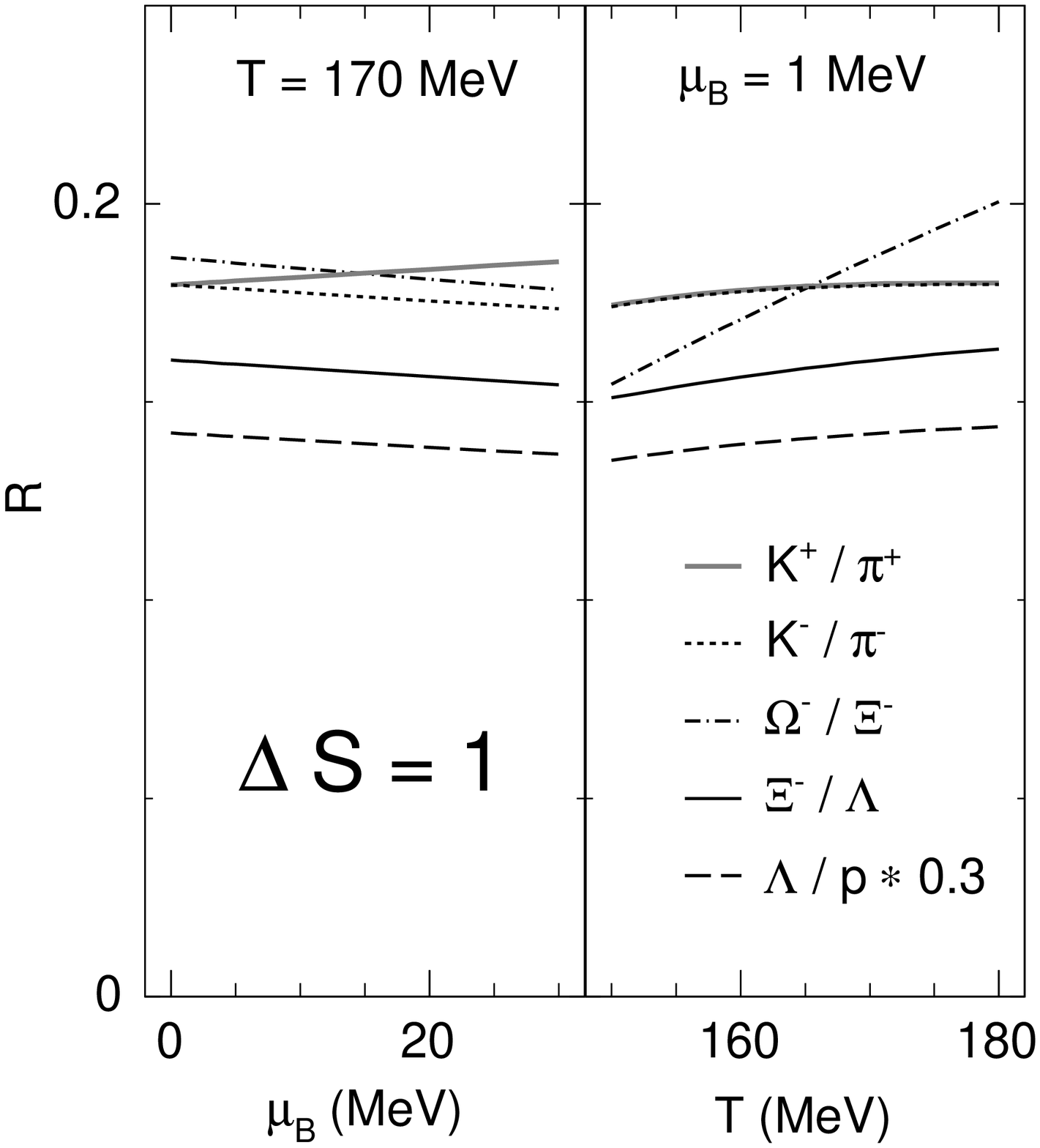}
\end{minipage}\hfill
\begin{minipage}[b]{0.05\linewidth}
~
\end{minipage}\hfill
\begin{minipage}[b]{16.5 cm}
\caption{\label{fig2ab}
Left: Particle ratios $R$ involving hyperons as a function of $T$ for $\mu_B$ = 1 MeV.
\newline
Right: Ratios $R$ of particles with unequal strangeness content as a function of $\mu_B$ for $T$ = 170 MeV and as a function of $T$ for $\mu_B$ = 1 MeV. The figures are taken from Ref.~\cite{lhc_pbpb}.}
\hfill
\end{minipage}
\end{center}
\end{figure}

\begin{table}
\begin{center}
\begin{minipage}[t]{16.5 cm}
\caption{Particle ratios in central Pb+Pb collisions at freeze-out conditions expected at the
					LHC: $T$ = (170$\pm$5) MeV  and $\mu_B$~=~${\rm 1^{+4}_{-1}}$ MeV. 
					The given errors correspond to the variation in the thermal parameters.
					Additional, systematic uncertainies in the ratios of the right column arise from
					unknown decay modes. They are smaller than 1\% in general, but reach 3\% in 
					the $\Xi^- / \Lambda$ ratio and 7\% in the  p$ / \pi^-$ and the $\Lambda / $p  
					ratios.}
\label{Table1}
\end{minipage}
\begin{tabular}{cc|cc}
\hline
\multicolumn{2}{c|}{ $\bar{h} / h$ Ratio } & \multicolumn{2}{c}{ mixed Ratio } \\ 
\hline
 $ \pi^+ / \pi^- ~ $ & ~ $ 0.9998^{+ 0.0002}_{- 0.0010}$ ~ & ~ $ \rm K^+ / \pi^+ ~ $ & ~ $ 0.180^{+ 0.001}_{- 0.001} $~  \\
 $ \rm K^+ / \rm K^- ~ $ & ~ $ 1.002^{+ 0.008}_{- 0.002}$ ~ & ~ $ \rm K^- / \pi^-~ $ & ~ $ 0.179^{+ 0.001}_{- 0.001} $~  \\
 $ \rm \bar{p} / \rm p	~ $ & ~ $ 0.989^{+ 0.011}_{- 0.045}$ ~ & ~ $ \rm p / \pi^-	~ $ & ~ $ 0.091^{+ 0.009}_{- 0.007} $~  \\
 $ \bar{\Lambda} / {\Lambda}~ $ & ~ $ 0.992^{+ 0.009}_{- 0.036}$ ~ &  ~ $ \Lambda / \rm p ~ $ & ~ $ 0.473^{+ 0.004}_{- 0.006} $~  \\
 $ \bar{\Xi}^+ / {\Xi}^-	~ $ & ~ $ 0.994^{+ 0.006}_{- 0.026}$ ~ &  ~ $ \Xi^- / \Lambda	~ $ & ~ $ 0.160^{+ 0.002}_{- 0.003} $~  \\
 $ \bar{\Omega}^+ / {\Omega}^-	~ $& ~ $ 0.997^{+ 0.003}_{- 0.015}$ ~ & ~ $ \Omega^- / \Xi^-~ $ & ~ $ 0.186^{+ 0.008}_{- 0.009} $~  \\
\hline
\end{tabular}
\end{center}
\end{table}
The knowledge of $T$ and $\mu_B$ allows to calculate any particle ratio, likewise, only a few ratios are sufficient to determine the thermal parameters experimentally and cross check the extrapolation. Some are particularly well suited: As one can see in Fig.~\ref{fig1ab} (right), generally antiparticle/particle ratios show a sensitivity to $\mu_B$, which is diluted with increasing strangeness content. The temperature dependence is very weak, hence the $\rm \bar{p}/p$ ratio can serve as a direct measure of the baryon chemical potential. Particle ratios with large mass difference are in particular sensitive to temperature changes as seen in Fig.~\ref{fig2ab}(left), we propose the $\Omega/\pi$ and $\rm \Omega/K$ ratio as thermometers.
%
%
The $\rm K/\pi$ ratio features a different property, Fig.~\ref{fig2ab} (right): It hardly changes in the $T$ and $\mu_B$ range under study. This precise prediction can be used to test the applicability of the statistical model at very high energies.

\section{Particle production in p+p interactions} \label{pp}
\subsection{\it Canonical description of strangeness conservation}
\begin{figure}[h]
\begin{center}
\begin{minipage}[b]{0.05\linewidth}
~
\end{minipage}\hfill
\begin{minipage}[b]{0.4\linewidth}
\centering
\includegraphics*[width=\linewidth]{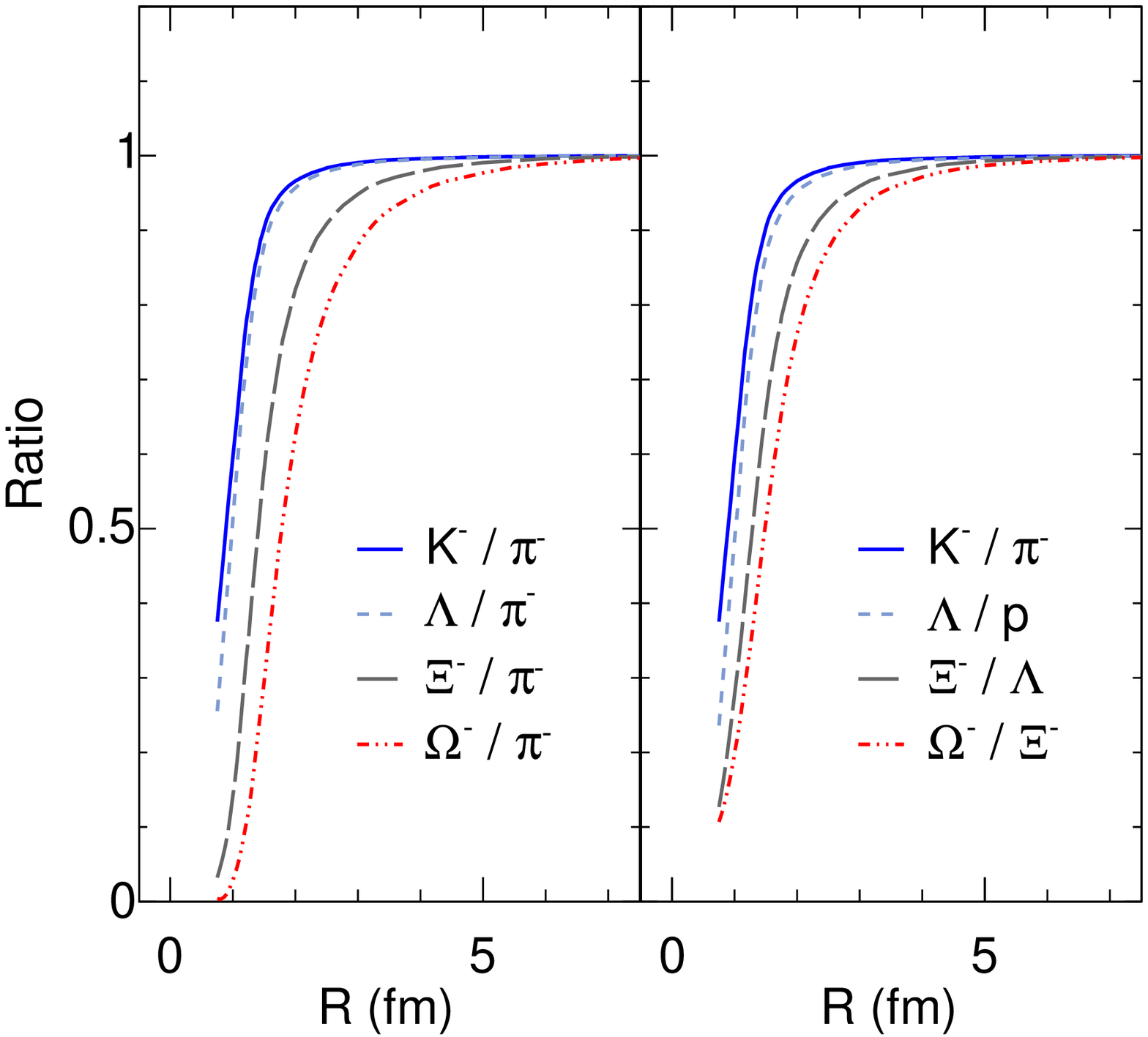}
\end{minipage}\hfill
\begin{minipage}[b]{0.05\linewidth}
~
\end{minipage}\hfill
\begin{minipage}[b]{0.4\linewidth}
\centering
\includegraphics*[width=\linewidth]{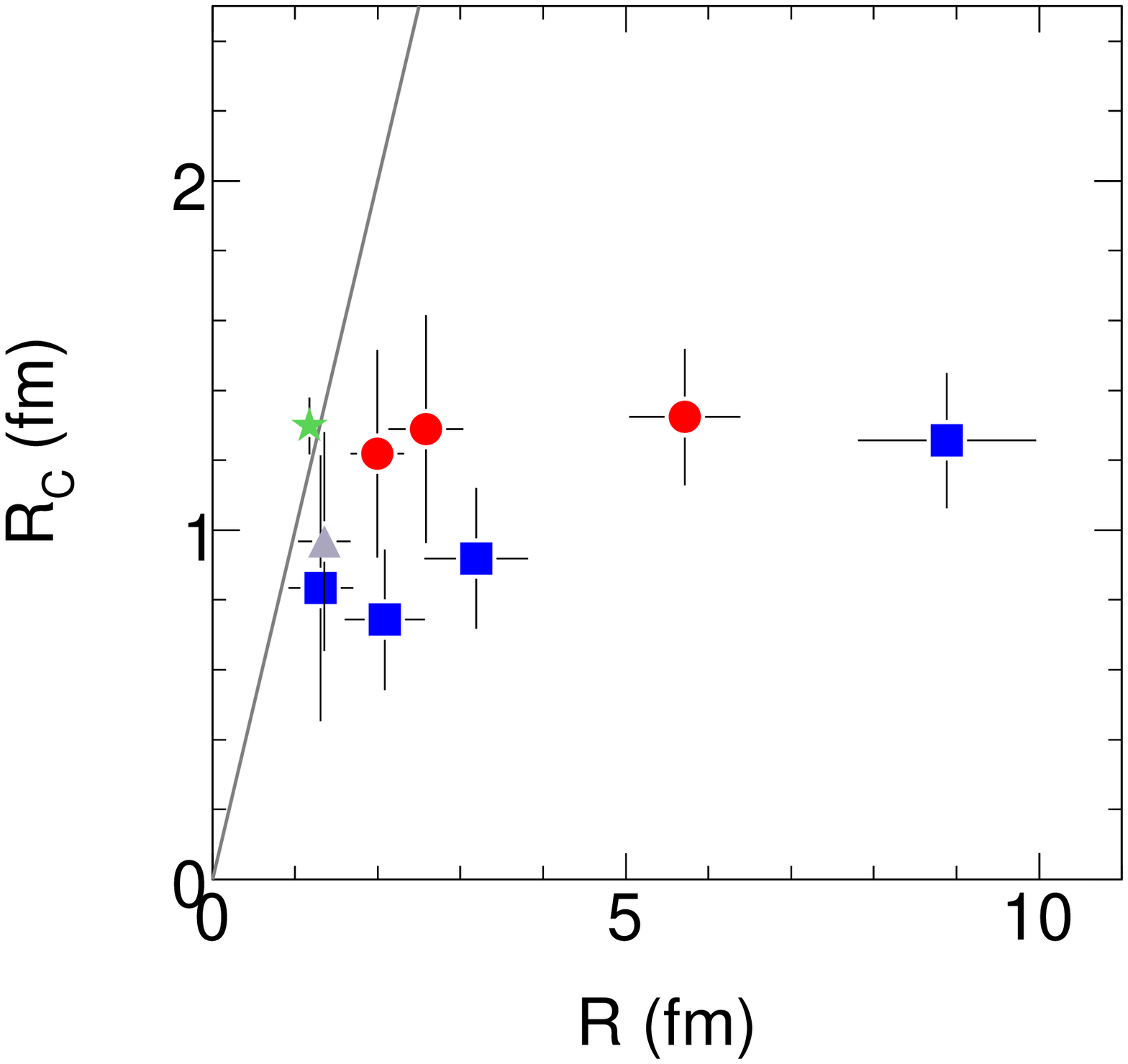}
\end{minipage}\hfill
\begin{minipage}[b]{0.05\linewidth}
~
\end{minipage}\hfill
\begin{minipage}[b]{16.5 cm}
\caption{\label{fig3ab}
Left: Different particle ratios as a function of the radius $R$ of a spherical volume. The temperature $T$ = 170 MeV and the baryon chemical potential $\mu_B$ = 255 MeV were chosen according to the thermal conditions at top SPS energy. All ratios are normalized to their  grand-canonical values. Taken from Ref.~\cite{syssize}.
\newline
Right: Cluster radius $R_C$ as a function of the fireball radius $R$ from fits to midrapidity densities (circles) and integrated yields (squares) from p+p and central C+C, Si+Si and Pb+Pb collisions at $\sqrt{s_{nn}}$ = 17 GeV from Ref.~\cite{syssize}. A larger data set of p+p interactions was analysed at the same energy (triangle) and at $\sqrt{s}$ = 200 GeV (star) in Ref.~\cite{pp_pred}. The full line indicates the proportionality $R_C = R$.
}
\hfill
\end{minipage}
\end{center}
\end{figure}

When dealing with particle production in p+p interactions, one has to allow for comparatively small abundance of strange hadrons. In general, if the number of particles carrying quantum numbers related to a conservation law is small, then the grand-canonical description no longer holds. In such a case conservation of charges has to be implemented exactly in the canonical ensemble. Here, we assume that charged particles and baryons are produced abundantly and consider charge and baryon number conservation to be fulfilled on the average in the grand-canonical ensemble. We refer only to local strangeness conservation. The density $n$ of strange particle $i$ carrying strangeness $s$ in the canonical ensemble $C$ can be obtained from its density in the grand-canonical ensemble $n_i$ by,
\begin{equation}
n_i^{C} \simeq n_i \frac{I_{s}(x)}{I_0(x)} \label{equ1}
\end{equation}
with the argument $x$ being a function of the volume and the sum over all particles carrying strangeness~$s$.
Figure~\ref{fig3ab} (left) illustrates this effect. Canonical suppression indeed depends on the strangeness quantum numbers of particles used in ratios and not only on their difference, as displayed in rhs of Fig.~\ref{fig3ab}~(left) and it becomes negligible if the system size is sufficiently large.
%

Strangeness suppression beyond the canonical suppression was found in the analysis of heavy-ion collisions at lower energies and in interactions of smaller systems~\cite{raf}. One solution to master this observation it to introduce an additional factor ($\gamma_S$) \cite{raf} which reduces the strange-particle phase-space but falls short of maintaining thermal equilibrium in the statistical model. Thereupon we adopt an alternative method, namely we assume that strangeness is conserved exactly in correlation volumes which can be smaller than the entire fireball. Requiring a partner for each strange quark within such a cluster leads to a strange-particle density according to the size of the correlation volume. Consequently, we employ two volume parameters, one that scales the grand-canonical particle densities and one that defines the degree of canonical suppression and thereby the abundance of strange particles.

Aiming at predictions for p+p interactions at LHC, we need knowledge of the energy dependence of strangeness production (and its canonical suppression in particular) and we have to be aware of the systems-size dependence of $T$ and $\mu_B$, which we have extrapolated to higher energies already. So we study collsions from p+p to Pb+Pb at the top SPS energy of $\sqrt{s_{nn}}$ = 17 GeV and we compare these results to an analysis of p+p data from RHIC at $\sqrt{s}$ = 200 GeV.

\begin{figure}[h]
\begin{center}
\begin{minipage}[b]{0.05\linewidth}
~
\end{minipage}\hfill
\begin{minipage}[b]{0.57\linewidth}
\centering
\includegraphics*[width=\linewidth]{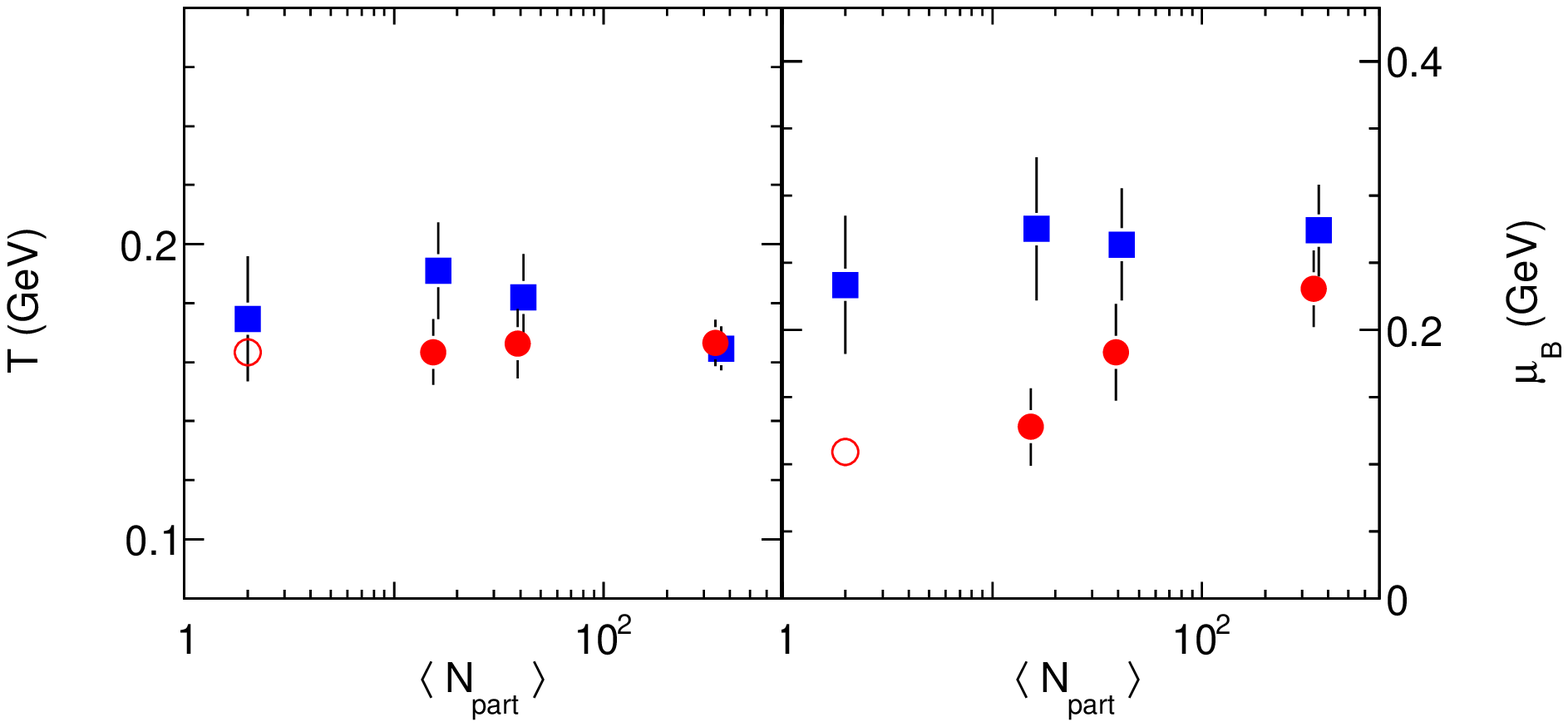}
\end{minipage}\hfill
\begin{minipage}[b]{0.05\linewidth}
~
\end{minipage}\hfill
\begin{minipage}[b]{0.27\linewidth}
\centering
\includegraphics*[width=\linewidth]{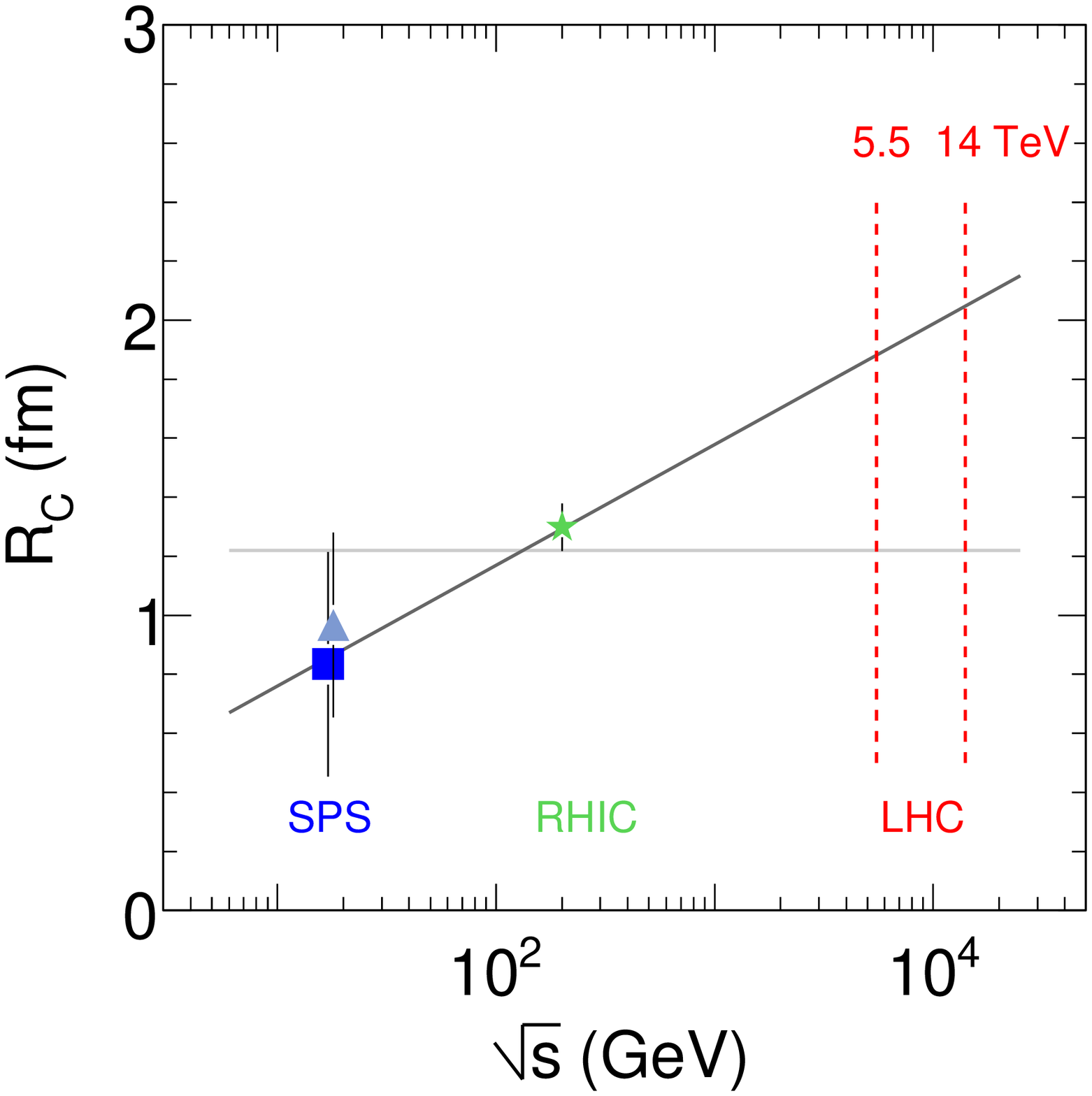}
\end{minipage}\hfill
\begin{minipage}[b]{0.05\linewidth}
~
\end{minipage}\hfill
\begin{minipage}[b]{16.5 cm}
\caption{\label{fig4ab}
Left: Chemical freeze-out temperature $T$ and baryon chemical potential $\mu_B$ from fits to midrapidity densities (circles) and integrated yields (squares) from p+p and central C+C, Si+Si and Pb+Pb collisions at $\sqrt{s_{nn}}$ = 17 GeV. The open circles refer to the baryon chemical potential at midrapidity extracted from the $\bar \Lambda/\Lambda$ ratio keeping $T$ as for C+C collisions. For details see Ref.~\cite{syssize}.
\newline
Right: Cluster radius $R_C$ as a function of energy, extracted from the analysis of SPS data (triangle) and RHIC data (star)~\cite{pp_pred}. Also shown is the  analysis~\cite{syssize} of a smaller set of SPS data (square). The lines illustrate possible evolutions towards LHC energies as discussed in the text.
}
\hfill
\end{minipage}
\end{center}
\end{figure}

Figure~\ref{fig3ab}(right) displays results on the cluster size: the correlated subvolmes are rather small and show a rather weak system size dependence, while the total volume increases significantly towards Pb+Pb, entailing that the clusters fill only in p+p interactions roughly the entire volume. Both, the temperature and the baryon chemical potential, exhibit no dependence on the size of the colliding nucleii as far as total multiplicities are considered, Fig.~\ref{fig4ab} (left). The smaller $\mu_B$ in midrapidity data can be attributed to less stopping.
%
%
\subsection{\it Extrapolation of thermal parameters and model predictions}
For predictions at LHC energies, we first present the extrapolation of the model parameters and then discuss the results.

The extrapolated temperature and baryon chemical potential for Pb+Pb collisions, based on Fig.~\ref{fig1ab} (left), was shown above. Since there is no significant systems-size dependence and we expect that stopping is an inferior and negligible effect at LHC and the potential will be anyhow very small in Pb+Pb collisions already, we use the same values of $T$ and $\mu_B$. In view of similar thermal conditions in p+p and Pb+Pb collisions, the essential difference in particle ratios in these systems will be due to canonical suppression. Thus, only particle ratios that contain the strange particles are different in elementary from heavy-ion collisions. In addition, the larger system size in Pb+Pb collisions at chemical freeze-out will clearly influence the total particle yields.

We have also analysed p+p interactions at $\sqrt{s}$~=~200~GeV (Fig.~\ref{fig3ab}, right) in order to extract the energy dependence of the strangeness correlation volume. There might be a trend for an increase with energy but within the errors a constant behaviour is possible, too. In Figure~\ref{fig4ab}, right, two extreme scenarios are indicated.

In the latter case, namely of an energy-independent cluster size, the initial size of the nucleons might regulate the strangeness production, and the strangeness suppression at LHC will be the same as at RHIC. Consequently, different ratios of strange/non-strange particle yields will be modified only  through the variation in $\mu_B$ which can be quantified by the Boltzmann factor $\exp(\pm\mu_B/T)$.
In the first case, of increasing subvolumes with $\sqrt{s}$, the hadron multiplicity in the final state might affect strangeness production, and the strangeness suppression at LHC will be weaker than at RHIC leading possibly to an equilibrated, canonical system without any additional suppression.
In this case, the cluster size should scale with the number of pions in the final state. This scenario could be verified experimentally at LHC by comparing the strange/non-strange particle ratios in p+p collisions for events with different pion multiplicities. If valid, then for sufficiently high pion multiplicity the strangeness production normalized to pion multiplicities might converge to the results expected in heavy-ion collisions. However, in view of the known data in elementary and heavy-ion collisions this is a very unlikely but not excluded scenario.
\begin{figure}[t]
\begin{center}
\begin{minipage}[b]{0.05\linewidth}
~
\end{minipage}\hfill
\begin{minipage}[b]{0.4\linewidth}
\centering
\includegraphics*[width=\linewidth]{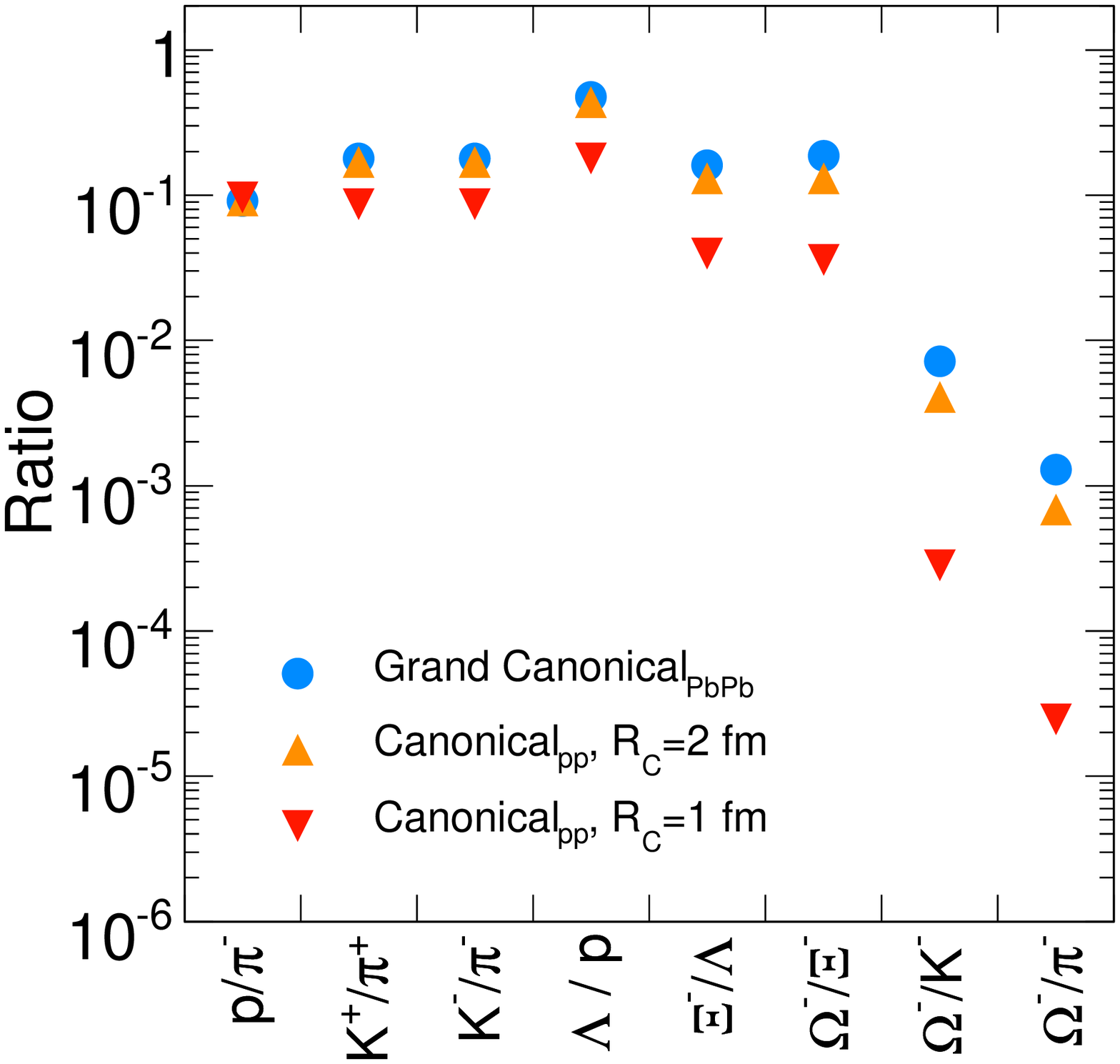}
\end{minipage}\hfill
\begin{minipage}[b]{0.05\linewidth}
~
\end{minipage}\hfill
\begin{minipage}[b]{0.4\linewidth}
\centering
\includegraphics*[width=\linewidth]{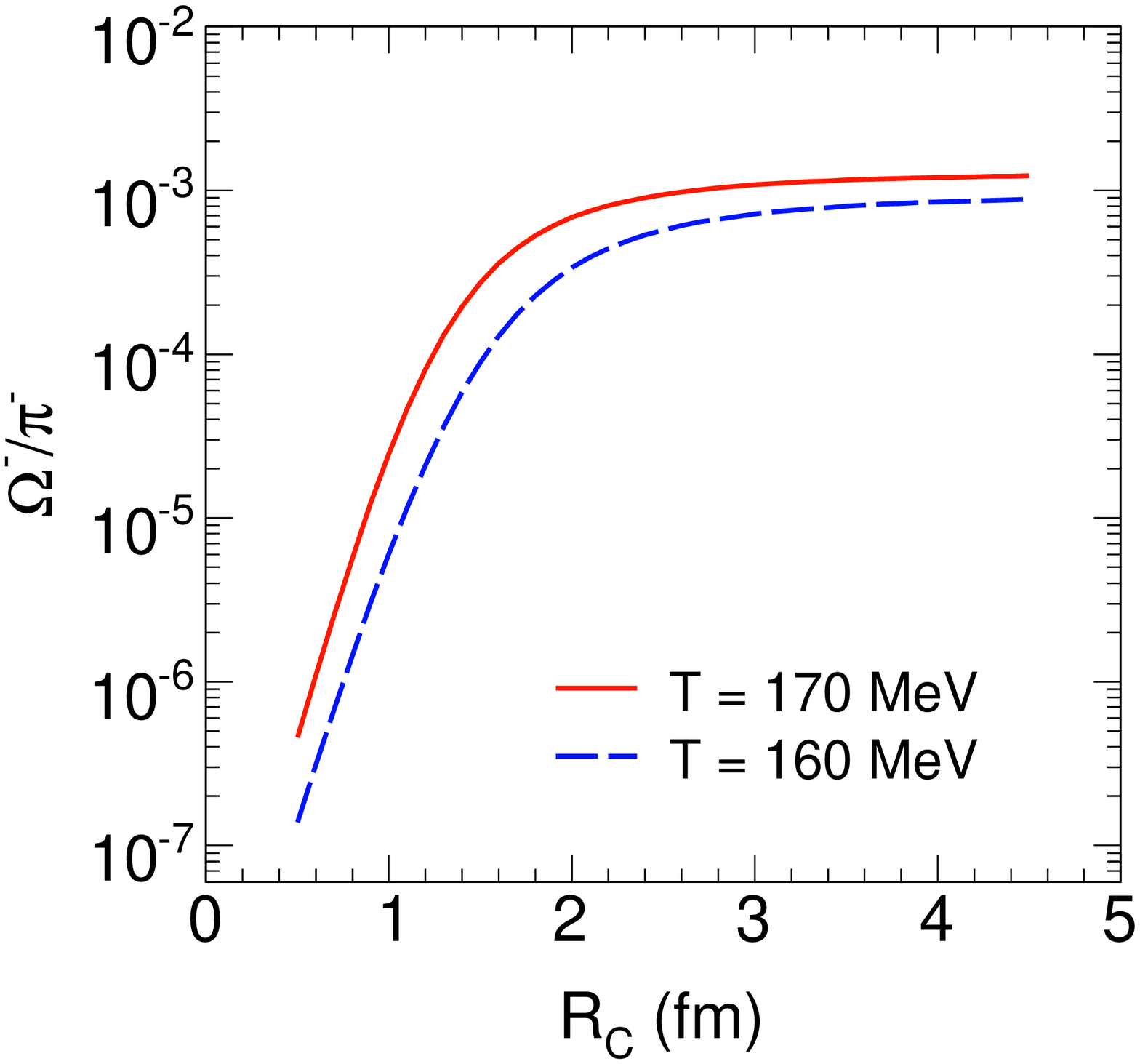}
\end{minipage}\hfill
\begin{minipage}[b]{0.05\linewidth}
~
\end{minipage}\hfill
\begin{minipage}[b]{16.5 cm}
\caption{\label{fig5ab}
Left: Predictions for various particle ratios using different values for the cluster size $R_C$.
\newline
Right: Particle ratio $\Omega/\pi$ as a function of the cluster size $R_C$ for two assumed temperatures, T = 160 MeV (dashed) and T = 170 MeV (full line).
}
\hfill
\end{minipage}
\end{center}
\end{figure}

%
%
\begin{table}
\begin{center}
\begin{minipage}[t]{16.5 cm}
\caption{Particle ratios in p+p interactions for thermal conditions expected 
at LHC. The extreme values of the cluster size $R_C$ span the band of expected 
numerical values. 
The grand-canonical values in the last column are included 
for comparison.}
\label{Table3}
\end{minipage}
\begin{tabular}{lccc}
\hline
Ratio & $R_C$ = 1 fm & $R_C$ = 2 fm & grand canon. \\
\hline
$p / \pi^-$ 		& 0.0970 	& 0.0920 & 0.0914 \\
$K^+ / \pi^+$ 		& 0.0871 	& 0.169 & 0.180 \\
$K^- / \pi^-$ 		& 0.0870 	& 0.169 & 0.179 \\
$\Lambda / p$ 		& 0.179 	& 0.436 & 0.473 \\
$\Xi^- / \Lambda$ 	& 0.0397 	& 0.130 & 0.160 \\
$\Omega^- / \Xi^-$ 	& 0.0358 	& 0.131 & 0.186 \\
$\Omega^- / K^-$ 	& 2.83 $\cdot ~ \rm 10^{-4}$ & 4.06 $\cdot ~ \rm 10^{-3}$ & 7.19 $\cdot ~ \rm 10^{-3}$ \\
$\Omega^- / \pi^-$ 	& 2.46 $\cdot ~ \rm 10^{-5}$ & 6.85 $\cdot ~ \rm 10^{-4}$ & 1.29 $\cdot ~ \rm 10^{-3}$ \\
\hline
\end{tabular}
\end{center}
\end{table}

Albeit the increase would be much weaker than linear, due to lack of data the energy dependence of the cluster size is unknown and a wide range is possible. Therefore the predictions for particle ratios are spread widely (Fig.~\ref{fig5ab}, left) and LHC data are essential to understand this behavior of strangeness suppression and its energy dependence. The same particle ratios as mentioned above for determine the temperature, namely $\Omega/\pi$ and $\Omega/K$, are well suited to quantify also the strangeness suppression, see Table~\ref{Table3}. This will cause no ambiguity because the volume dependence (Fig.~\ref{fig5ab}, right) is much stronger than the variation with temperature, displayed in Fig.~\ref{fig2ab} (left).
%
\section{Conclusion} \label{summary}
Within the statistical thermal model, we provide predictions for particle production in Pb+Pb collisions at LHC energy with freeze-out parameters obtained from the extrapolation of existing results to higher energies.
Also the sensitivities of various ratios with respect to the temperature and the baryon chemical potential are discussed and we have shown that  the $\rm \bar{p} / \rm p$ ratio is the best suited observable to extract the value of the baryon chemical potential at chemical freeze-out.  The $\Omega^- / \pi^-$ and the $\Omega^-$/K$^-$ ratio are proposed as thermometers to extract experimentally the chemical freeze-out temperature in central Pb+Pb collisions at LHC.

The situation is different for p+p interactions since one has to allow for possible strangeness suppression. We have discussed the role and the influence of strangeness correlation volume on particle production. Existing data constrain extrapolations only to a rather broad band of predictions.
However, the $\Omega/\pi$ and $\rm \Omega/K$ ratios are excellent probes of strangeness correlations and/or strangeness suppression mechanism in p+p collisions. Furthermore, comparing strangeness production at the LHC in events with different pion multiplicities can provide a deep insights into our understanding of strangeness suppression from A+A to p+p collisions.
%
%

\end{document}